\documentstyle[12pt]{article}
\begin{document}

\thispagestyle{empty}
\begin{flushright}
JHU-TIPAC-94021\\
BONN-TH-94-27\\
hepth@xxx/9412032\\
December 1994
\end{flushright}

\bigskip\bigskip\begin{center} {\bf
\Large{Manifest supersymmetry and the ADHM construction of instantons}}
\end{center} \vskip 1.0truecm

\centerline{\bf A. Galperin}
\vskip5mm
\centerline{Department of Physics and Astronomy}
\centerline{Johns Hopkins University}
\centerline{Baltimore, MD 21218, USA}
\vskip5mm
\centerline{\bf and}
\vskip5mm
\centerline{\bf E. Sokatchev}
\vskip5mm
\centerline{Physikalishes Institut}
\centerline{Universit\"at Bonn}
\centerline{Nu{\ss}allee 12, 53115 Bonn, Germany}
\vskip5mm

\bigskip \nopagebreak \begin{abstract}
\noindent
We present the $(0,4)$ superspace version of Witten's sigma model
construction for ADHM instantons. We use the harmonic superspace
formalism, which exploits the three complex structures common to both
$(0,4)$ supersymmetry and self-dual Yang-Mills theory. A novel feature of
the superspace formulation is the manifest interplay between the ADHM
construction and its twistor counterpart.\end{abstract}

\newpage\setcounter{page}1

\section{Introduction}

In a recent very interesting paper \cite{Witten} Witten constructed a
$(0,4)$ supersymmetric linear sigma model in two dimensions with a
potential term with surprising properties. Under certain assumptions about the
structure of the multiplets involved, the Yukawa couplings of the model
were found to satisfy the ADHM equations of the instantons construction
\cite{Adhm}. In the infrared regime the massless chiral fermions in the
model turned out to be coupled to the instanton gauge field, obtained
exactly in accord with the ADHM prescription.  The fact that $(0,4)$
supersymmetry implies self-duality of the target-space gauge fields is not
new (see, for instance, \cite{Howe}). However, it is a very unexpected
feature of Witten's model that it requires couplings to {\it instantons},
i.e., self-dual gauge fields with finite action.

Non-linear $(0,4)$ sigma models with potentials had been discussed before
(see, e.g., \cite{Hull},\cite{George}) and the restrictions on the
potential imposed by supersymmetry had been found. However, this had been
done in the traditional way of \cite{Alvarez}. There one starts with one
supersymmetry and then tries to add three more. This approach breaks the
intrinsic automorphism $SO(4) \sim SU(2)\times SU(2)'$ of $(0,4)$
supersymmetry to $SO(3)$. Witten carried out the analysis in a fully
$SO(4)$ covariant way, which greatly facilitated not only finding the
conditions on the potential but also solving them. In addition, Witten
required that the potential be invariant under one of the two $SU(2)$
automorphism groups. All of this allowed him to discover the remarkable
relation between this type of sigma models and the ADHM construction.

In \cite{Witten} the sigma model was given in terms of component fields and on
shell, i.e., without manifest $(0,4)$ supersymmetry. In this paper we
present the superspace version of Witten's model. The natural setting for
$N=4$ off-shell supersymmetry is $SU(2)$ harmonic superspace \cite{harms}. On
the one hand, it turns out that one of the multiplets used by Witten (the
chiral fermion one) can, in general, only exist off shell with an infinite set
of
auxiliary fields. Harmonic superfields are just the right objects of this
type. On the other hand, the harmonic approach is best adapted to exploit
the three complex structures common for self-dual Yang-Mills (SDYM) fields
and $(0,4)$ supersymmetry. In this way we find a natural bridge between
the ADHM construction for instantons and its twistor counterpart given by
Ward \cite{Ward}.

In section \ref{sec1} we first recall some basic facts about $SU(2)$
harmonics and show how they are used \cite{harW} to reformulate the
twistor approach of Ward \cite{Ward} to SDYM theory. We also give a brief
summary of the ADHM construction in a notation adapted to field theory. In
section \ref{04} we give the harmonic superspace description of the
relevant $(0,4)$ off-shell multiplets, namely, the scalar, chiral fermion
and ``twisted" scalar multiplet and construct their coupling of potential
type. It turns out that the ``twisted" scalar multiplet of \cite{Witten}
is described by an {\it abelian gauge odd} $(0,4)$ superfield (its gauge
invariance is an artifact of supersymmetry and
is not related to the Yang-Mills group). It is precisely this gauge
invariance that severely restricts the possible couplings and leads to the
ADHM conditions. The component version of the ``twisted" multiplet is
obtained in a Wess-Zumino (not manifestly supersymmetric) gauge for
this superfield. However, there also exists a manifestly supersymmetric
gauge. In it part of the chiral fermion superfields are gauged away,
another part are absorbed by the ``twisted" multiplet superfields, which
become massive. The remaining massless chiral fermion superfields interact
with a target-space SDYM field, which naturally arises in its twistor
(harmonic) form. Thus the superspace sigma model allows us to see the
relation between the ADHM construction and its twistor
counterpart. It also makes it possible to generalize the linear
sigma model to non-linear ones, i.e., to switch on a hyper-K\"ahler
background, again given in its twistor version.

In Appendix A we explain why the chiral fermion multiplet requires, in
general, infinitely many auxiliary fields off shell. In Appendix B we show
how the $(0,4)$ superfields used in this paper
can be obtained from $N=2,D=4$ ones by trivial
dimensional reduction to $(4,4)$ supersymmetry in two dimensions
and further truncation  down to $(0,4)$.
The set of $N=2,D=4$ harmonic superfields is very simple, it just consists of
matter (hypermultiplet) and abelian gauge ones \cite{harms}.

\section{Self-duality, $(0,4)$ supersymmetry and $SU(2)$
harmonics}\label{sec1}

In this section we shall recall the basic concepts of the method of
harmonic space \cite{harms}. We shall do this in two different contexts,
that of SDYM theory and of $(0,4)$ supersymmetry. One of our tasks will be
to convince the reader that the harmonic method allows one to exhibit very
clearly the remarkable similarity between these two theories. The basic
reason for this similarity is the existence of three complex structures in
both cases.

\subsection{Self-dual Yang-Mills theory}

Let us start with self-duality. SDYM fields can be defined in
a Euclidean space of dimension $4k$, $R^{4k}$. This space has, in general,
$SO(4k)$ as a symmetry group. In the context of self-duality it is convenient
to reduce this symmetry to $Sp(1)\times Sp(k)$ and to introduce coordinates
$X^{AY}$. Here $A=1,2$ is an $SU(2)\sim Sp(1)$ index and $Y=1,\ldots,2k$ is an
$Sp(k)$ index. The coordinates are real in the sense $\overline{X^{AY}}
= \epsilon_{AB}\epsilon_{YZ}X^{BZ}$, where $\epsilon$ are the
antisymmetric constant tensors of the two symplectic groups. \footnote{We
use the following convention to raise/lower symplectic indices: $X_A=
\epsilon_{AB}X^B, \; X^A= \epsilon^{AB}X_B; \;
\epsilon^{AB}\epsilon_{BC}=\delta^A_C$. We always assume summation over
repeated indices.}

The Yang-Mills fields for some internal symmetry group are introduced via the
covariant derivatives $\nabla_{AY} = \partial_{AY} + A_{AY}(X)$. Their
commutator
defines the field strength tensor, which is decomposed under $Sp(1)\times
Sp(k)$ as follows:
\begin{equation}\label{fs}
[\nabla_{AY},\nabla_{BZ}] = \epsilon_{AB} F_{(YZ)} + F_{(AB)[YZ]}\; .
\end{equation}
Here $F_{(AB)[YZ]}$ is symmetric in $A,B$ and antisymmetric in $Y,Z$.
Self-duality means to impose the condition
\begin{equation}\label{sd}
F_{(AB)[YZ]} = 0\; .
\end{equation}
These first-order equations imply the standard  Yang-Mills equations
$\nabla^{AY} F_{AYBZ} = 0$, but unlike in the latter case, one is able to find
exact solutions to  eqs. (\ref{sd}).

\subsection{Harmonic variables}\label{sdym}

The way to the exact solutions of the self-duality equations (\ref{sd})
goes, in one way or another, through finding an interpretation of
(\ref{sd}) as an integrability condition. We shall do that in the harmonic
framework \cite{harW}.  We must point out that the harmonic treatment of
self-duality as integrability is a modification of the twistor one given
by Ward \cite{Ward}. We believe that the harmonic approach might be
somewhat closer to physicists with a field-theoretical background.

The harmonic variables are defined as $SU(2)$ matrices:
\begin{equation}\label{hv}
u^{\pm A} \in SU(2): \ \ u^{+A}u^-_A=1, \ \ u^-_A= \overline {u^{+A}}\; .
\end{equation}
In what follows we shall not use any particular parametrization of
$SU(2)$, but instead we shall use the matrices (\ref{hv}) themselves as
(constrained) variables. One can write down three differential operators
compatible with the definition (\ref{hv}):
\begin{equation}\label{hd}
D^{++}=u^{+A}{\partial\over\partial u^{-A}}, \ \ D^{--} = \overline{D^{++}},
\ \ D^0 =u^{+A}{\partial\over\partial u^{+A}} -
u^{-A}{\partial\over\partial u^{-A}}\; .
\end{equation}
The third one $D^0$ clearly counts the $U(1)$ charge $\pm$ of the
variables $u^{\pm A}$.
We shall define {\it harmonic functions} as eigenfunctions of this charge
operator,
\begin{equation}\label{ch}
D^0f^q(u) = q f^q(u), \;\;\;\; q =0, \pm 1, \pm 2, \ldots,
\end{equation}
that have a {\it harmonic
expansion}
\begin{equation}\label{he}
f^q(u) = \sum^{\infty}_{n=0} f^{A_1\ldots A_{n+q}B_1\ldots B_n} u^+_{(A_1}
\ldots u^+_{A_{n+q}}u^-_{B_1}\ldots u^-_{B_n)}
\end{equation}
(for $q\geq 0$; for $q<0$ the expression is analogous).
In fact, this is nothing but the spherical harmonic expansion of a (square
integrable) function (or tensor, if $q\neq 0$) on the sphere
$S^2\sim SU(2)/U(1)$. The coset $SU(2)/U(1)$ is obtained
via the homogeneity condition (\ref{ch}). We want to emphasize that this way
of describing the sphere (as opposed to using, e.g., a complex
parametrization) is global, we don't have to worry about the analytic
properties of our functions in different coordinate patches.

We should mention that for even values of the $U(1)$ charge $q$ the
harmonic functions (\ref{he}) can be made real with respect to the
following special conjugation:
\begin{equation}\label{conj}
\widetilde{f^{AB\ldots}} = \overline{f^{AB\ldots}},\ \ \
\widetilde{u^{\pm A}} = u^\pm_A\; , \;\;\;\;
\widetilde{u^\pm_A} = - u^{\pm A}\; .
\end{equation}
In fact this is just a combination of usual complex conjugation and of the
antipodal map on $S^2$, realized in terms of $u^\pm$ (see \cite{harms} for
details). This conjugation will systematically be used below.

To complete our short introduction into the harmonic calculus, here are two
formulas which we shall use in what follows. A harmonic integral is defined
by the simple rule
\begin{equation}\label{integ}
\int du\; 1 = 1, \ \ \int du\; u^+_{(A_1}
\ldots u^+_{A_{p}}u^-_{B_1}\ldots u^-_{B_q)} = 0 \ \mbox{for $p$ and/or
$q>0$}\; .
\end{equation}
In other words, the harmonic integral projects out the singlet part in the
expansion of the integrand. The above integration rule is compatible with
integration by parts for the derivatives (\ref{hd}). In addition, one can
prove the following identity
\begin{equation}\label{df}
D^{++}_1{1\over u^+_1u^+_2} = \delta^{+,-}(u_1,u_2)\; ,
\end{equation}
where $u^+_1u^+_2 \equiv u^{+A}_1u^+_{2A}$ and $\delta^{+,-}(u_1,u_2)$ is a
harmonic delta function. \footnote{This identity is equivalent to
$\partial/\partial \bar z (1/z) = \pi\delta(z)$. Indeed, using a complex
parametrization of $S^2$, one can show that $D^{++}\sim
\partial/\partial\bar z$ and $u^+_1u^+_2 \sim z_1 - z_2$.}

Now let us come back to the self-duality condition (\ref{sd}). Multiplying eq.
(\ref{fs}) by $u^{+A}u^{+B}$ and introducing the notation
\begin{equation}\label{n+}
\nabla^+_Y = u^{+A}\nabla_{AY}\; ,
\end{equation}
we can rewrite (\ref{sd}) as follows
\begin{equation}\label{sd+}
[\nabla^+_Y, \nabla^+_Z] = 0\; .
\end{equation}
Note also that from the definition (\ref{n+}) and from the properties of the
harmonic derivative $D^{++}$ immediately follows that
\begin{equation}\label{+++}
[D^{++}, \nabla^+_Y] = 0\; .
\end{equation}
The important point is that the two new conditions (\ref{sd+}) and
(\ref{+++}) are in fact {\it equivalent} to the original self-duality one
(\ref{sd}). Indeed, let the connection $A^+_Y(x,u)$ in $\nabla^+_Y$ be a
harmonic function satisfying (\ref{+++}). Inspecting the harmonic
expansion (\ref{he}), we easily see that the solution of (\ref{+++}) must
have the form (\ref{n+}). Then, using the fact that $u^+_A,u^+_B$ are {\it
arbitrary} commuting doublets, from (\ref{sd+}) we derive (\ref{sd}).

Equation (\ref{sd+}) provides the desired integrability interpretation of the
self-duality condition. Namely, consider a matter field $\phi(X,u)$ satisfying
the following {\it analyticity} condition
\begin{equation}\label{an}
\nabla^+_Y \phi(X,u) = 0
\end{equation}
in a Yang-Mills background. Then eq. (\ref{sd+}) is obviously the
integrability condition for the existence of such fields. Further, eq.
(\ref{sd+}) has the following (local) general solution of ``pure gauge" type
\begin{equation}\label{pg}
A^+_Y = h^{-1}\partial^+_Y h\; ,
\end{equation}
where $h(X,u)$ is some {\it harmonic dependent} element of the gauge group.
This new object allows us to rewrite the analyticity condition (\ref{an}) in
a much simpler way
\begin{equation}\label{flan}
\partial^+_Y\Phi = 0, \ \ \ \Phi \equiv h\phi\; ,
\end{equation}
which has solutions in the form of functions holomorphic
with respect to the projected variable $X^{+Y}= u^+_{A} X^{AY}$:
\begin{equation}\label{sol}
\partial^+_Y\Phi = 0 \ \ \Rightarrow \ \ \Phi = \Phi(X^+,u)\; .
\end{equation}
In other words, the ``pure gauge"  transformation (\ref{pg}), which generates
the solutions to (\ref{sd+}), allows us to define a new gauge frame, in which
holomorphicity (\ref{sol}) becomes manifest. It is clear that this procedure
can only be non-trivial because of the harmonic dependence of the fields.

The last point about the harmonic interpretation of self-duality concerns the
harmonic derivative $D^{++}$. The change of gauge frame $\phi \rightarrow
h\phi$ made the connection in the derivative $\nabla^+_Y$ disappear.
However, at the same time the previously ``short" derivative $D^{++}$ acquires
now a connection:
\begin{equation}\label{hco'}
D^{++} \ \ \rightarrow \ \ \nabla^{++} = D^{++} + V^{++}\; .
\end{equation}
where
\begin{equation}\label{hco}
V^{++} = hD^{++}h^{-1}\; .
\end{equation}
In these new terms the second self-duality constraint (\ref{+++}) becomes
\begin{equation}\label{+++'}
[\nabla^{++}, \partial^+_Y]=0 \ \ \Rightarrow \ \ V^{++}=V^{++}(X^+,u)\; .
\end{equation}
The conclusion is that the solutions of the SDYM equations can be
parametrized (locally) by the holomorphic harmonic connection $V^{++}$
from (\ref{hco'}),(\ref{+++'}). We should mention here that the harmonic
connection $V^{++}$ is in a sense the analog of the ``twistor transform"
of the SDYM field in the approach of Ward \cite{Ward}.  Given an (almost)
arbitrary $V^{++}(X^+,u)$, one is in principle able to find the ``pure
gauge" transformation matrix $h(X,u)$ from (\ref{hco}) and then
reconstruct the Yang-Mills connection $A$ from (\ref{pg}). In practice,
however, this amounts to solving the non-linear differential equation
(\ref{hco}) on $S^2$, which is not an easy task. Moreover, in this
framework it is not obvious which $V^{++}$ would generate solutions of the
SDYM equations with finite action ({\it instantons}).

\subsection{The ADHM construction}\label{quaternion}

An alternative way
of obtaining (or, rather, parametrizing) the instanton solutions is the
ADHM construction, which is the main subject of this article.  One of the
results of this paper will be to establish a one-to-one correspondence
(modulo gauge freedom) between
the instanton-generating connections $V^{++}$ and the building blocks of the
ADHM construction (see section \ref{oleg}). It turns out that the
superspace realization of Witten's sigma model makes this
correspondence transparent. It should be mentioned that
a similar relation between $V^{++}$ and the ADHM construction
 has already been presented in a different
context in \cite{Oleg}.

For the reader's convenience we give here a short summary of the ADHM
construction. More details can be found in the original instanton literature
(\cite{Adhm},\cite{ErWe},\cite{Corr}). Here we shall treat only the case of an
$SO(n)$ gauge group. The principal difference from the standard presentation of
the ADHM construction will be that we do not use the traditional quaternionic
notation for the various matrices. The obvious reason for this is that we want
to construct a field theory (following Witten), which reproduces the ADHM
construction. Using quaternion-valued fields in an action is rather
inconvenient.

The starting point of the ADHM construction for the case $SO(n)$ is the
rectangular matrix $\Delta^a_{AY'}$ with indices $a=1,\ldots,n+4k'$,
$Y'=1,\ldots,2k'$ ($n$ and $k'$ are any two positive integers); $A$ is the
same $Sp(1)$ index as above.  This matrix is required to satisfy
several conditions:
\begin{itemize}
\item It must be real in the sense
\begin{equation}\label{adhm0}
\overline{\Delta^a_{AY'} } = \epsilon^{AB}\epsilon^{Y'Z'}\Delta^a_{BZ'}\; .
\end{equation}
\item It must be linear in $X$,
\begin{equation}\label{adhm1}
\Delta^a_{AY'} = \alpha^a_{AY'}  + \beta^a_{Y'Y} X^{Y}_A
\end{equation}
with constant matrices $\alpha,\beta$.
\item It must satisfy the algebraic constraint
\begin{equation}\label{ADHM2}
\Delta^a_{AY'}\Delta^a_{BZ'} = \epsilon_{AB} R_{Y'Z'}\; ,
\end{equation}
where $R_{Y'Z'}$ is an invertible antisymmetric $2k'\times 2k'$ matrix.
\item The matrices $\alpha,\beta$ must have maximal rank.
\end{itemize}

To obtain the SDYM field one needs another real rectangular matrix $v^a_i$,
where
the index $i=1,\ldots,n$ is an $SO(n)$ one. It is defined to be orthogonal to
the matrix $\Delta$,
\begin{equation}\label{orgo}
v^a_i\Delta^a_{AY'} = 0
\end{equation}
and orthonormal,
\begin{equation}\label{orno}
v^a_iv^a_j=\delta_{ij}\; .
\end{equation}
Then the $SO(n)$ gauge field is given by the simple expression
\begin{equation}\label{siex}
(A_{AY})_{ij} =v^a_i\partial_{AY}v^a_j\; .
\end{equation}
It is not hard to check that (\ref{siex}) is indeed a self-dual field in the
sense (\ref{sd}). In the process one essentially uses the linear dependence of
$\Delta$ on $X$. It is not much more difficult to show that (\ref{siex})
corresponds to an {\it instanton} solution with finite action and instanton
number $k'$. What is really hard is the proof that this construction gives {\it
all instanton solutions}. For more details see
\cite{Adhm},\cite{ErWe},\cite{Corr}.

{}From the procedure described above it is clear that the matrices $\Delta$ and
$v$ contain a considerable arbitrariness. Indeed, the resulting gauge field
(\ref{siex}) is not affected by global $SO(n+4k')$ transformations on the index
$a$. Further, the index $Y'$ can be multiplied by global $GL(2k',C)$ matrices,
compatible with the reality condition (\ref{adhm0}), which reduces them to
$GL(k',Q)$ ones. Of course, local $SO(n)$
transformations on the index $i$ correspond to the usual Yang-Mills freedom in
the gauge field. The $SO(n+4k')\times GL(k',Q) $ arbitrariness is
essential when one  counts the
number of independent parameters of the instanton solution. In particular, one
is able to fix a very
simple canonical form for the coefficient matrices $\alpha,\beta$
(\ref{adhm1}).  In
the orthogonal case and for $R^4$ ($k=1$) it is \cite{ErWe}:
\begin{equation}\label{canon}
\alpha^a_{AY'} \ \rightarrow \ \left(\begin{array}{c} b_{4k'\times n}
\\ B_{4k'\times 4k'}
\end{array}\right), \ \ \
\beta^a_{YY'} \ \rightarrow \ \left(\begin{array}{c} 0_{4k'\times n} \\
1_{4k'\times 4k'}
\end{array} \right).
\end{equation}
The remaining non-trivial matrices $b, B$ satisfy algebraic
constraints following from (\ref{ADHM2}). This, together with the rest of
the symmetry transformations leads to the correct number of independent
parameters in the instanton solutions \cite{ErWe},\cite{Corr}.

The point about the freedom of redefinitions above
is rather important. As we shall see in section \ref{04}, in the sigma model
construction of Witten one is forced to interpret $Y'$ as an $Sp(k')$ index.
The consequences of this will be discussed later on.

Finally, for those who are more familiar with the traditional quaternionic
notation in the instanton literature, we indicate the correspondence.
\footnote{We are grateful to F. Delduc for helping us establish this
correspondence.} In the four-dimensional case ($k=1$) \footnote{An
analogous treatment of the case $k>1$ has been given in \cite{CGK}.}
$X^{AY}$ is a $2\times 2$ matrix, $X^{AY}=X^\mu(\sigma_\mu)^{AY}$.
The matrix $\Delta$ can be rewritten as a quaternionic
one if one splits the index $Y'\rightarrow A'y'$, where $A'$ is a new $Sp(1)$
index and now $y'=1,\ldots,k'$. Then the two $Sp(1)$ indices $AA'$ can be
saturated with Pauli matrices and $\Delta$ becomes $\Delta^a_{y'}$ with
quaternionic entries. The analog of condition (\ref{ADHM2}) is \cite{ErWe}
\begin{equation}\label{eric}
(\Delta^a_{y'})^{\dagger} q \Delta^a_{z'} = \mbox{Re}(R_{y'z'}q)\; ,
\end{equation}
valid for an arbitrary quaternion $q$.

\subsection{$(0,4)$ supersymmetry}

The harmonic variables are very useful in application to $(0,4)$ supersymmetry
on a two-dimensional world sheet.
The basic reason is the presence of three complex structures in this
supersymmetry.

The $(0,4)$ world sheet can be regarded as a superspace with coordinates
$x_{++}, x_{--}, \theta^{AA'}_+$. Here $\pm$ indicate the Lorentz
($SO(1,1)$) weights (to avoid confusion with the harmonic $U(1)$ charges
we shall always write the weights as lower and the charges as upper
indices). The Grassmann variables $\theta^{AA'}_+$ carry doublet indices
of the $SO(4)\sim SU(2)\times SU(2)'$ automorphism group of $(0,4)$
supersymmetry and satisfy the reality condition $\overline{\theta^{AA'}} =
\epsilon_{AB}\epsilon_{A'B'} \theta^{BB'}$. The spinor covariant derivatives
(the counterparts of the supersymmetry generators)
\begin{equation}\label{D}
D_{-AA'} = {\partial\over\partial\theta^{AA'}_+} + i\theta_{+AA'}
{\partial\over\partial x_{++}}
\end{equation}
satisfy the $(0,4)$ supersymmetry algebra
\begin{equation}\label{susy}
\{ D_{-AA'}, D_{-BB'}\} = 2i\epsilon_{AB}\epsilon_{A'B'} \partial_{--}\; .
\end{equation}
Comparing eq. (\ref{susy}) with (\ref{fs}) and (\ref{sd}),
we see the first similarity between
$(0,4)$ supersymmetry and SDYM theory in $R^{4k}$. This suggests to
apply the same trick of projecting the $SU(2)$ indices $A,B$ with harmonic
variables, as we did in (\ref{n+}). Defining
\begin{equation}\label{D+}
D^+_{-A'} \equiv u^{+A}D_{-AA'}\; ,
\end{equation}
we can rewrite (\ref{susy}) as follows
\begin{equation}\label{++}
\{ D^+_{-A'}, D^+_{-B'}\} = 0\; .
\end{equation}
Note that after this projection the torsion term from the right-hand side of
(\ref{susy}) disappeared, just like the curvature term in (\ref{sd+}). This
analogy goes even further. In the Yang-Mills case we interpreted eq.
(\ref{sd+}) as the integrability condition for the existence of analytic
fields defined by (\ref{an}). Here we can define {\it Grassmann analytic
superfields} satisfying
\begin{equation}\label{ga}
D^+_{-A'}\Phi(x,\theta,u) = 0\; ,
\end{equation}
with eq. (\ref{++}) as the integrability condition. Further, in subsection
\ref{sdym} we proceeded by changing the gauge frame as in
(\ref{pg}),(\ref{flan}), in order to ``shorten" the covariant derivative
$\nabla^+_Y$. Here the analogous operation is the introduction of a new
{\it analytic basis} in harmonic (i.e., extended by adding the harmonic
variables) superspace
\begin{equation}\label{bas}
\hat x_{++} = x_{++} + i\theta^{AA'}_+\theta^{B}_{+A'} u^+_{(A}u^-_{B)},
\ \ x_{--}, \ \ \theta^{\pm A'}_{+}= u^\pm_A \theta^{AA'}_+, \ \ u^\pm\; .
\end{equation}
Its existence is guaranteed by the integrability conditions (\ref{++}).
In it the derivative $D^+_{-A'}$ is just the partial one: $D^+_{-A'} =
\partial/\partial\theta^{-A'}_+$.  Then the Grassmann analyticity
condition (\ref{ga}) can be solved in the form
\begin{equation}\label{asf}
D^+_{-A'}\Phi(x,\theta,u) = 0 \ \ \Rightarrow \ \ \Phi = \Phi(\hat x_{++},
x_{--}, \theta^+_+, u)\; .
\end{equation}
At the same time the harmonic derivative $D^{++}$, which acquired a
connection in the new frame in subsection \ref{sdym}, here gets a vielbein
term
\begin{equation}\label{anD++}
D^{++} = u^{+A}{\partial\over\partial u^{-A}} + i\theta^{+A'}_+\theta^+_{+A'}
{\partial\over\partial\hat x_{++}}\; .
\end{equation}
To complete the parallel between the two pictures, note that the analog of
(\ref{+++}) here is
\begin{equation}\label{hgr}
[D^{++}, D^+_{-A'}] = 0\; ,
\end{equation}
which is basis-independent.

The analytic superfields defined in
(\ref{asf}) can have a non-vanishing $U(1)$ harmonic charge, $\Phi^q
(x,\theta^+,u)$. \footnote{To simplify the notation we shall not write
explicitly $\hat x_{++}$ when it is clear that we work in the analytic
basis (\ref{bas}).} They have a very short Grassmann expansion,
\begin{equation}\label{expa}
\Phi^q(x,\theta^+,u) = \phi^q(x,u) + \theta^{+A'}_+\xi^{q-1}_{-A'}(x,u)
+(\theta^+_+)^2 f^{q-2}_{--}(x,u)\; ,
\end{equation}
where $(\theta^+_+)^2 \equiv \theta^{+A'}_+\theta^+_{+A'}\;$.  The coefficients
in (\ref{expa}) are harmonic-dependent fields (note that the overall
$U(1)$ charge $q$ is conserved in all the terms in (\ref{expa})).
We should also mention that in certain cases the harmonic analytic
superfields (\ref{expa}) can be made
real in the sense of the special conjugation (\ref{conj}).

The careful reader may note that we have ``harmonized" just one  $SU(2)$
subgroup of the $SO(4)$ automorphism group, but haven't touched
the other one, $SU(2)'$.  It is this fact that makes
harmonic superfields very useful in CFT. There, $SU(2)'$ is a part of the
$(0,4)$ superconformal group, while $SU(2)$ is not.
In the model considered below, the action is $SU(2)'$ invariant and not
$SU(2)$ invariant. So, it has the right symmetries to flow in the infrared to
an $(0,4)$ CFT.

\section{The $(0,4)$ sigma model}\label{04}

In this section we give the superspace formulation of the sigma model used
by Witten in \cite{Witten} to show the relation between $(0,4)$
supersymmetry and the ADHM constructions of instantons. One of the
superfields needed turns out to have an abelian gauge invariance (which
has nothing to do with the gauge symmetry of the Yang-Mills theory). It is
this invariance which, to a large extent, is responsible for the peculiar
coupling leading to the ADHM construction. We show that the manifestly
supersymmetric sigma model action gives rise to the twistor transform of the
instanton gauge field. We also
point out a subtlety concerning the exact relationship between
the $(0,4)$ sigma model and the ADHM construction.

\subsection{The free supermultiplets}

Witten makes use of three types of $(0,4)$ supermultiplets. Here we shall
give their superspace counterparts. Their origin is perhaps more clear if they
are regarded as obtained by trivial dimensional reduction
and truncation from the two basic
$N=2, D=4$ superfields, namely,
the matter (hypermultiplet) and the {\it abelian} gauge multiplet. We shall
explain this in Appendix B.

The first multiplet includes the coordinates
$X^{AY}$ of the Euclidean target space $R^{4k}$, in which the Yang-Mills fields
will be defined. In
harmonic superspace it is described by the analytic superfields (cf.
(\ref{expa}))
\begin{equation}\label{X}
X^{+Y}(x,\theta^+,u) = X^{+Y}(x,u) + i\theta^{+A'}_+\psi^Y_{-A'}(x,u)
+(\theta^+_+)^2 f^{-Y}_{--}(x,u)\; .
\end{equation}
These superfields can be chosen real in the sense of the conjugation
defined in (\ref{conj}):
\begin{equation}\label{Xreal}
\widetilde{X^{+Y}} = \epsilon_{YZ} X^{+Z}\; .
\end{equation}
Each field in (\ref{X}) is harmonic, i.e., has an infinite harmonic expansion
on $S^2$. However, the multiplet we are looking for should have a finite
number of physical fields. It turns out that we can truncate the harmonic
expansions in (\ref{X}) by imposing the following harmonic irreducibility
condition
\begin{equation}\label{irr}
D^{++} X^{+Y} = 0\; .
\end{equation}
It is not hard to obtain the component solution of this constraint. Using the
analytic basis form of $D^{++}$ (\ref{anD++}), for $X^{+Y}(x,u)$ we find
($\partial^{++}$ denotes the purely harmonic part of (\ref{anD++}))
\begin{equation}\label{eqX}
\partial^{++} X^{+Y}(x,u) = 0 \ \ \Rightarrow \ \ X^{+Y}(x,u) =
u^+_A X^{AY}(x)\; ,
\end{equation}
as follows from the harmonic expansion (\ref{he}) for $q=+1$. Similarly, for
the other two fields in (\ref{X}) we get
\begin{equation}\label{solu}
\psi^Y_{-A'}(x,u) = \psi^Y_{-A'}(x), \ \ \ f^{-Y}_{--}(x,u) = -i u^-_A
\partial_{--}X^{AY}(x)\; .
\end{equation}
The component fields are real because of (\ref{Xreal}),
$\overline{X^{AY}}=
\epsilon_{AB}\epsilon_{YZ}X^{BZ}, \; \overline{\psi^{A'Y}}=
\epsilon_{A'B'}\epsilon_{YZ}\psi^{B'Z}$.
So, as a result of the constraint (\ref{irr}) the harmonic dependence of
the fields in (\ref{X}) was reduced to a linear one. Since the constraint
(\ref{irr}) is manifestly supersymmetric and does not involve equations of
motion for the component fields, we conclude that the fields in
(\ref{eqX}),(\ref{solu}) form an off-shell $(0,4)$ supersymmetry
multiplet. It coincides with the one given in \cite{Witten}.

The action for the above multiplet is given as an integral over the analytic
superspace:
\begin{equation}\label{acX}
S_X = i\int d^2x du d^2\theta^+_+ \; X^{+Y}\partial_{++} X^+_Y\; .
\end{equation}
Note that in constructing (\ref{acX}) we took special care to match the
$U(1)$ charge $-2$ and Lorentz weight $-2$ of the Grassmann measure with
those of the integrand, so that the action is both an $SU(2)$ and Lorentz
singlet. To obtain the component content of (\ref{acX}) we substitute the
short form (\ref{eqX}),(\ref{solu}) of the expansion (\ref{X}) in
(\ref{acX}) and perform two more steps. First, we do the Grassmann
integral, i.e., we pick out only the $(\theta^+_+)^2$ terms. Then we do
the harmonic integral according to the rules (\ref{integ}), which amounts
to extracting the $SU(2)$ singlet part. The result is
\begin{equation}\label{comX}
S_X = \int d^2x\; \left( X^{AY}\partial_{++}\partial_{--} X_{AY} +
{i\over 2}\psi^{A'Y}_-\partial_{++}
\psi_{-A'Y}\right)\; .
\end{equation}
This is the action for $4k$ real scalars $X^{AY}$ and their chiral spinor
superpartners $\psi^{A'Y}_-$. The multiplet in (\ref{comX}) is clearly free
and massless, which corresponds to a flat target space $R^{4k}$. The way
to endow the target space with a non-flat (necessarily hyper-K\"ahler)
metric or
even a torsion has been studied in \cite{hypK},\cite{hypKK}. One simply
replaces the flat irreducibility condition (\ref{irr}) by
\begin{equation}\label{curv}
D^{++} X^{+Y} = {\cal L}^{+3Y}(X^+,u)\; ,
\end{equation}
with an arbitrary function of $U(1)$ charge $+3$ (the hyper-K\"ahler case,
i.e., no torsion, corresponds to choosing ${\cal L}^{+3Y} =
(\partial/\partial X^+_Y) {\cal L}^{+4}$, see \cite{hypK} for more
details). In what follows we shall not be interested in such
generalizations, we prefer to keep (\ref{irr}) flat.

The second type of $(0,4)$ supermultiplet used by Witten involves only chiral
fermions (at least on shell) of chirality, opposite to that in (\ref{comX}).
Its harmonic superspace description is very simple indeed. Take the following
{\it anticommuting} and real (in the sense of (\ref{conj})) superfields
\begin{equation}\label{Lam}
\Lambda^a_+(x,\theta^+,u) = \lambda^a_+(x,u) + \theta^{+A'}_+g^{-a}_{A'}(x,u)
+i(\theta^+_+)^2  \sigma^{--a}_{-}(x,u)\; .
\end{equation}
The action for them is
\begin{equation}\label{acL}
S_\Lambda = {1\over 2}\int d^2x du d^2\theta^+_+ \;
\Lambda^{a}_+D^{++} \Lambda^a_{+}\; .
\end{equation}
{}From here it is clear that we can regard the external index
$a=1,\ldots,n+4k'$ as an $SO(n+4k')$ one: integrating $D^{++}$ in
(\ref{acL}) by parts changes the sign, but the superfields $\Lambda^a_+$
anticommute, so the trace $aa$ is symmetric.

Obtaining the component content of (\ref{acL}) is as easy as in the case of
(\ref{acX}). First we do the Grassmann integral and find
\begin{equation}\label{comL}
S_\Lambda = \int d^2xdu \; \left( {i\over 2}\lambda^{a}_+\partial_{--}
\lambda^a_{+}
+ i\sigma^{--a}_{-} \partial^{++}\lambda^{a}_+ + {1\over 4}
 g^{-aA'} \partial^{++} g^{-a}_{A'} \right)\; .
\end{equation}
Note that the space-time derivative $\partial_{--}$ in (\ref{comL})
originates from the harmonic derivative (\ref{anD++}). The field
$\sigma^{--a}_{-}$ serves as a Lagrange multiplier for the harmonic
condition $\partial^{++}\lambda^a_+(x,u) =0$ which makes $\lambda^a_+$
harmonic independent.  The harmonic-dependent field $g^{-aA'}(x,u)$ is
clearly auxiliary too. Eliminating both auxiliary harmonic fields, we
obtain simply the action for $n+4k'$ free chiral fermions
\begin{equation}\label{chfe}
S_\Lambda = {i\over 2}\int d^2x \;
\lambda^{a}_+(x)\partial_{--}\lambda^a_{+}(x)\; .
\end{equation}
This represents an on-shell supermultiplet, in which supersymmetry acts in a
trivial way. However, and we want to stress on this, the off-shell version
of the multiplet (\ref{comL}) requires an {\it infinite number} of
auxiliary fields. Assuming a {\it finite} number of auxiliary fields, one
can show by the standard ``no-go'' counting arguments of \cite{RocekSiegel}
that the number of chiral fermions must be an integer multiple of four.
Moreover, for the multiplets with a finite number of auxiliary fields,
the natural $SO(n+4k')$ symmetry of the free action (\ref{chfe})
is necessarily broken (see Appendix A).
 Naively, one might conclude that it is
impossible to have an arbitrary number of chiral fermions in an off-shell
representation of $(0,4)$ supersymmetry \cite{Gates}. However, the
harmonic formalism
avoids this by using an infinity of auxiliary fields. This is an analog of
the $N=2$ hypermultiplet \cite{harms}
and $N=3$ supersymmetric Yang-Mills multiplet \cite{N=3} in four dimensions,
for which there exist no finite off-shell multiplets.

Finally, Witten utilizes the so-called ``twisted" scalar multiplet, in
which the $SU(2)$ indices carried by the bosons and fermions are
interchanged (as compared to the standard multiplet (\ref{comX})). Its
superspace description turns out to be quite unusual. This time we need
 a set of {\it anticommuting abelian gauge} superfields
\begin{equation}\label{Phi}
\Phi^{+Y'}_+(x,\theta^+,u) = \rho^{+Y'}_+(x,u) +
\theta^{+A'}_+\phi^{Y'}_{A'}(x,u) +i(\theta^+_+)^2  \chi^{-Y'}_{-}(x,u)\; ,
\end{equation}
satisfying the reality condition
$\widetilde{\Phi^{+Y'}_+}=\epsilon_{Y'Z'}\Phi^{+Z'}_+$ (see (\ref{conj})).
Here $Y'=1,\ldots,2k'$ is an index of some new symplectic group $Sp(k')$, as
we shall see below (it turns out that $k'$ is just the number of instantons in
the corresponding ADHM construction).
These superfields undergo the following abelian gauge transformations
\begin{equation}\label{gau}
\delta \Phi^{+Y'}_+ = D^{++} \omega^{-Y'}_+
\end{equation}
with analytic parameters $\omega^{-Y'}_+(x,\theta^+,u)$. We want to
stress that this gauge invariance has nothing to do with the target space
gauge group.  It is an artifact of the superspace description of the
multiplet.

The r\^ole of the gauge invariance (\ref{gau}) may become more transparent
if we make comparison with an abelian gauge field in four dimensions,
$A_\mu(x)$.  It is known that this field contains two representations of
the Poincar\'e group with spins 1 and 0. There are two ways to eliminate
the spin 0 part.  One is to impose an irreducibility condition, e.g.,
$\partial^\mu A_\mu =0$, the other is to submit $A_\mu$ to gauge
transformations $\delta A_\mu = \partial_\mu
\omega(x)$ and write down a gauge invariant action.
Something similar we observe
here. The analog of the irreducibility condition $\partial^\mu A_\mu =0$
is eq. (\ref{irr}) for the superfield $X^+$. The effect of this condition
is that the superfield becomes short in the harmonic sense, which means
the elimination of an (infinite number) of extra degrees of freedom. The
analog of the second, gauge mechanism for $A_\mu$ is given by (\ref{gau}).
To understand what happens we should look at the expansion of the
parameter
\begin{equation}\label{om}
\omega^{-Y'}_+(x,\theta^+,u) = \tau^{-Y'}_+(x,u) + \theta^{+A'}_+
l^{--Y'}_{A'}(x,u)
+i(\theta^+_+)^2  \mu^{-3Y'}_{-}(x,u)\; .
\end{equation}
{}From (\ref{Phi}) and (\ref{om}) and using (\ref{anD++}), we get, for
instance, $\delta\rho^+_+(x,u) = \partial^{++}\tau^-_+(x,u)$. Comparing
the harmonic expansions (\ref{he}) of $\rho^+$ and $\tau^-$, we easily see
that each term in the expansion of $\rho^+$ has its counterpart in that of
$\tau^-$. So, the component $\rho^+$ can be completely gauged away.
Similar arguments show that the expansions of the parameters $l^{--}$ and
$\mu^{-3}$ are a little ``shorter" than those of the fields $\phi$ and
$\chi^{-}$, correspondingly. What is missing is just the lowest order
term, the smallest $SU(2)$ representations in each expansion. Thus, the
fields $\phi^{Y'}_{A'}(x,u)$ and $\chi^{-Y'}_{-}(x,u)$ can also be gauged
away, except for the first terms in their harmonic expansions, the fields
$\phi^{Y'}_{A'}(x)$ and $u^-_A\chi^{Y'A}_{-}(x)$. The net result of all
this is the following ``short", i.e., irreducible harmonic superfield in
the {\it Wess-Zumino-type gauge}
\begin{equation}\label{WZ}
\Phi^{+Y'}_+(x,\theta^+,u) = \theta^{+A'}_+\phi^{Y'}_{A'}(x)
+i(\theta^+_+)^2  u^{-}_A \chi^{Y'A}_{-}(x)\; .
\end{equation}
This is precisely the content of the ``twisted" multiplet of Witten. We note
that this multiplet is off shell, just like the one described by the
superfield $X^+$. In fact, in terms of component fields the difference
between the two multiplets is very small, just the two automorphism groups
$SU(2)$ are flipped. However, we have seen that their superspace
descriptions are quite different, and the reason for this is clear: we have
harmonic variables for one of the $SU(2)$ groups only. We would also like to
stress that in the
Wess-Zumino gauge (\ref{WZ}) one loses manifest supersymmetry.

Next, we have to find a gauge invariant action for the superfield $\Phi^+_+$.
It is modeled after the action for a gauge superfield for $N=2,D=4$
supersymmetry \cite{harms} and has the form
\begin{equation}\label{acP}
S_\Phi =i \int d^2x d^4\theta_+ du_1du_2\; {1 \over u^+_1u^+_2}
\Phi^{+Y'}_+(1)\partial_{++} \Phi^+_{+Y'}(2)\; .
\end{equation}
The notation $\Phi^+_+(1) $ means that the analytic superfield is written
down in an analytic basis (\ref{bas}) defined by the harmonic variable
$u_1$; similarly, $\Phi^+_+(2) $ is given in another basis, defined by
$u_2$.  Since both superfields appear in the same integral, they should be
written down in the same non-analytic basis $x_{\pm\pm},
\theta^{AA'}_+,u$. This explains
why the Grassmann integral in (\ref{acP}) is taken over the full superspace
and not over an analytic subspace, as in (\ref{acX}) or (\ref{acL}). Note
that, as always, we take care of matching $U(1)$ charges and Lorentz
weights (this accounts for the presence of $\partial_{++}$ in (\ref{acP})).
Note also that the contraction ${}^{Y'}{}_{Y'} \equiv {}^{Y'}\epsilon_{Y'Z'}
{}^{Z'}$ must be of the type $Sp(k')$. Indeed, interchanging 1 and 2 involves
integration by parts of $\partial_{++}$, flipping the odd superfields
$\Phi(1)$, $\Phi(2)$ and using $u^+_1u^+_2 = - u^+_2u^+_1$, so the fourth
antisymmetric factor $\epsilon_{Y'Z'}$ restores the symmetry required by
the double harmonic integral. The issue of the symmetry of this kinetic term is
important and we shall come back to it later on.

The reason for the exotic form of (\ref{acP}) is the gauge invariance
(\ref{gau}). It works in the following way. Varying in (\ref{acP}) with
respect to, e.g., $\Phi^+_+(1) $ and according to the transformation law
(\ref{gau}) makes the harmonic derivative $D^{++}_1$ appear under the
integral. Integrating it by parts, we see that it only acts upon the
harmonic distribution $(u^+_1u^+_2)^{-1}$ (since $\Phi^+_+(2) $ depends on
the second harmonic variable $u_2$). Then we use the formula (\ref{df})
and obtain a harmonic delta function, which removes one of the harmonic
integrals. After this both superfields $\Phi^+_+$ become analytic with
respect to the same basis, i.e., they depend only on the two Grassmann
variables $\theta^{+A'}_+$. However, the Grassmann integral in (\ref{acP})
is over $d^4\theta$, so it gives 0. This establishes the gauge invariance
of the action (\ref{acP}).

The action (\ref{acP}) may seem strange because of its
harmonic non-locality, but it is only apparent. Indeed,
in the Wess-Zumino gauge (\ref{WZ}) all the fields have short harmonic
expansions. Working out the Grassmann integral (not forgetting the different
arguments of $\Phi(1,2)$), we obtain factors of $(u^+_1u^+_2)$ in the
numerator
which cancel out the singular denominator. Then the double harmonic integral
becomes trivial and we find
\begin{equation}\label{comP}
S_\Phi = \int d^2x\; \left( \phi^{Y'A'}\partial_{++}\partial_{--}
\phi_{Y'A'} + {i\over 2}\chi^{Y'A}_-
\partial_{++} \chi_{-Y'A} \right)\; ,
\end{equation}
which is the ``twisted" multiplet action used by Witten.

\subsection{Interaction and the ADHM construction}\label{flat}

The main question now is how to couple the above multiplets. Following the
idea of Witten, we are only interested in couplings of the potential type,
i.e., without space-time derivatives. The coupling should be controlled by
a parameter $m$ with the dimension of mass. The aim is to examine the
resulting theory in the limit $m\rightarrow\infty$ and show that it flows
into a CFT. The latter are known to have an unbroken $SU(2)$ invariance,
and this is put as a requirement in Witten's construction. In our harmonic
superspace description this corresponds to preserving the symmetry
$SU(2)'$ (indices $A'$).

The interaction terms of the action can be integrals over either the full
$(0,4)$ harmonic superspace $\int d^2x du d^4\theta$ or the analytic
superspace $\int d^2x du d^2\theta^+_+ $; they should involve a positive
power of $m$. To see how this can be arranged, let us first examine the
dimensions of our superfields $X^{+Y}$, $\Lambda_+^a$ and $\Phi^{+Y'}_+$.
The position of the physical spinors (dimension 1/2) $\psi_-$ in
(\ref{X}), $\lambda_+$ in (\ref{Lam}) and $\chi_-$ in (\ref{Phi}) fixes
the dimensions $[X]=0$, $[\Lambda] = 1/2, [\Phi] = -1/2$.
It is easy to see that the full superspace integrals are ruled out. Indeed,
the full measure has dimension $0$ and Lorentz weight $0$. The presence of
the mass in $m\int d^2x d^4\theta$ requires at least a pair of $\Phi^+_+$
superfields
($[\Phi^2]=-1$), but this is not  consistent with the Lorentz invariance.
Thus, we are left with analytic superspace
interaction terms only. The analytic measure $\int d^2x du
d^2\theta^+_+ $ has dimension 1, Lorentz weight $-2$ and $U(1)$ charge
$-2$. There are only two possible coupling terms in the action, in
which the dimensions, charges and weights are matched (we
omit the indices):
\begin{eqnarray}
S_{\Phi\Lambda}&=&m\int d^2x du d^2\theta^+_+\; \Phi^+_+\Lambda_+ v^+(X^+,
u)\; ,
\label{m} \\
S_{\Phi\Lambda}&=&m^2\int d^2x du d^2\theta^+_+\;  \Phi^+_+\Phi^+_+ t(X^+,
u)\; .
\label{mm}
\end{eqnarray}
Here $v^+$ and $t$ are arbitrary functions of the dimensionless and
weightless superfields $X^+$ and of the harmonic variables $u$.

Let us first investigate the term (\ref{m}). To get an
idea how it can be constructed, we first take a simplified case, in which
$n=0$ and the index $a$ can be written as a pair of symplectic indices, $a
= AY'$. Then the charged object $v^+$ can be the harmonic variable $u^+_A$
itself. Thus we come to the following coupling term
\begin{equation}\label{coup}
S_{\Phi\Lambda} =m \int d^2x du d^2\theta^+_+ \;  \Phi^{+Y'}_+ u^{+A}
\Lambda_{+AY'}\; .
\end{equation}
The most important point now is to make sure that the
coupling (\ref{coup}) is invariant under the
gauge transformation (\ref{gau}). It is very easy to see that to this end the
superfield $\Lambda_{+AY'}$ must also transform as follows
\begin{equation}\label{gauL}
\delta \Lambda_{+AY'} =m u^+_A \omega^-_{+Y'}\; .
\end{equation}
Indeed, the variation with respect to $\Lambda$ in the kinetic term
(\ref{acL}) compensates for that of $\Phi$ in (\ref{coup}), whereas
$\delta \Lambda$ in (\ref{coup}) is annihilated by the harmonic $u^+$
$(u^{+A}u^+_A = 0)$.

The second possibility (\ref{mm}) to construct a coupling term is now
clearly ruled out by the gauge invariance (\ref{gau}) (however, soon we
shall see a term of the type (\ref{mm}) appearing in a {\it fixed gauge}).

Surprisingly enough, in spite of the presence of gauge invariance, the
coupling (\ref{coup}) is nothing but a mass term. There are two ways to
see this. One is to examine the component Lagrangian, the other is to do it
directly in terms of superfields. We postpone the former until we come to the
general interaction and do the latter, which is in fact much easier. Let us
decompose the index $A$ of $\Lambda_{+AY'}$ in the $u^{\pm A}$ basis:
\begin{equation}\label{deco}
\Lambda_{+AY'} = u^+_A \Lambda^-_{+Y'} + u^-_A \Lambda^+_{+Y'}\; , \;\;\;
\Lambda^-_{+Y'} \equiv -u^{-A}\Lambda_{+AY'}\; , \;\;
\Lambda^+_{+Y'} \equiv u^{+A}\Lambda_{+AY'}\; .
\end{equation}
Then from (\ref{gauL}) we see that $\delta \Lambda^-_{+Y'} =m
\omega^-_{+Y'}$,
which allows us to fix the following {\it supersymmetric gauge} (as opposed to
the non-supersymmetric Wess-Zumino gauge (\ref{WZ}))
\begin{equation}\label{sugau}
\Lambda^-_{+Y'}  = 0\; .
\end{equation}
Substituting this into (\ref{coup}) and the kinetic term (\ref{acL}), we
obtain two action terms involving the remaining projection $\Lambda^+_{+Y'} $:
\begin{equation}\label{alg}
\int d^2x du d^2\theta^+_+ \; \left(-{1\over 2}
\Lambda^{+Y'}_{+}\Lambda^+_{+Y'} +
m \Phi^{+Y'}_+ \Lambda^+_{+Y'}\right)\; .
\end{equation}
Clearly, $\Lambda^+_{+Y'}$ can be eliminated from (\ref{alg}) by means of its
algebraic field equation $\Lambda^+_{+Y'} = m \Phi^+_{+Y'}$. Then we are left
only with the superfield $\Phi$ which has the action (recall (\ref{acP}))
\begin{equation}\label{massac}
S_\Phi =i \int d^2x d^4\theta_+ du_1du_2\; {1 \over u^+_1u^+_2}
\Phi^{+Y'}_+(1)\partial_{++} \Phi^+_{+Y'}(2) + {m^2\over 2} \int d^2x du
d^2\theta^+_+ \; \Phi^{+Y'}_+\Phi^+_{+Y'}\; .
\end{equation}
The very appearance of (\ref{massac}) shows that it is the action for $2k'$
massive superfields. We conclude that when the number of the multiplets
$\Lambda^a$ equals $4k'$, the coupling term (\ref{coup}) is in fact a mass
term for the $4k'$ real bosons $\phi^{A'Y'}$ contained in $\Phi$ and for
the $4k'$ pairs of right-handed fermions $\chi^{AY'}_-$ from $\Phi$ and
left-handed ones $\lambda^{AY'}_+$ from $\Lambda$.

Let us now come back to the general case when $n \neq 0$. We can try to
generalize the coupling (\ref{coup}) by replacing $u^+_A$ by a matrix
$v^{+a}_{Y'}(X^+,u)$, satisfying the reality condition $\widetilde{v^{+a}_{Y'}}
= \epsilon^{Y'Z'}v^{+a}_{Z'}$. It cannot depend on the superfields $\Phi$ or
$\Lambda$
because of Lorentz invariance, but it can still be a function of the
weightless superfields $X^{+Y}$ and of the harmonic variables. So, we
write down
\begin{equation}\label{int}
S_{int} = m \int d^2x du d^2\theta^+_+ \;  \Phi^{+Y'}_+ v^{+a}_{Y'}(X^+,u)
\Lambda_+^a\; .
\end{equation}

The main question is how to make (\ref{int}) compatible with the gauge
invariance (\ref{gau}). Like in (\ref{gauL}), we can try
\begin{equation}\label{gauLv}
\delta \Lambda^a_{+} = mv^{+a}_{Y'}(X^+,u)  \omega^{-Y'}_+\; .
\end{equation}
It is not hard to see that this transformation compensates for
$\delta\Phi$ in (\ref{int}), provided the following two conditions hold:
\begin{equation}\label{c1}
v^{+a}_{Y'}v^{+a}_{Z'} = 0\; ,
\end{equation}
\begin{equation}\label{c2}
D^{++} v^{+a}_{Y'}(X^+,u) = 0\; .
\end{equation}

Condition (\ref{c2}) is very restrictive. Indeed, assume that the
function $v^{+a}_{Y'}(X^+,u)$ is regular, i.e., can be expanded in a Taylor
series in $X^+$:
\begin{equation}\label{tay}
v^{+a}_{Y'}(X^+,u) = \sum^{\infty}_{p=0}v^{(1-p)a}_{Y'Y_1\ldots Y_p}(u)
X^{+Y_1} \ldots X^{+Y_p}\; .
\end{equation}
Since $X^+$ satisfies the irreducibility condition (\ref{irr}), (\ref{c2})
implies $D^{++}v^{(1-p)a}_{Y'Y_1\ldots Y_p}(u) = 0$, where $1-p$ is the $U(1)$
charge. The only non-vanishing solution to
this is (see the harmonic expansion (\ref{he}))
\begin{equation}\label{lin}
v^{+a}_{Y'}(X^+,u) = u^{+A} \alpha^a_{AY'} + \beta^a_{Y'Y} X^{+Y}\; ,
\end{equation}
where the matrices $\alpha,\beta$ are constant. In other words, the matrix
$v^{+a}_{Y'}(X^+,u)$ can at most depend linearly on $X^+$. If we put
$\theta^+=0$, i.e., consider only the lowest components in the superfield
$X^+$ (see (\ref{eqX})), then (\ref{lin}) becomes
\begin{equation}\label{lin'}
v^{+a}_{Y'}(X^+,u)|_{\theta=0} = u^{+A} (\alpha^a_{AY'}  +
\beta^a_{Y'Y} X^{Y}_A) \equiv u^{+A}
\Delta^a_{AY'}\; .
\end{equation}

The other condition (\ref{c1}) is purely algebraic. Putting (\ref{lin'})
in it and removing the harmonic variables $u^{+A}u^{+B}$, we get
\begin{equation}\label{ADHM}
\Delta^a_{(AY'}\Delta^a_{B)Z'} = 0, \;\; \mbox{i.e.,} \;\;\;\;
\Delta^a_{AY'}\Delta^a_{BZ'} = \epsilon_{AB} R_{Y'Z'}\; .
\end{equation}
What we have obtained are precisely the two defining conditions
(\ref{adhm1}),(\ref{ADHM2}) on the matrix
$\Delta$, used as a starting point in the ADHM construction for instantons. As
we mentioned in subsection \ref{quaternion}, in the ADHM construction one
requires that the
matrices $\alpha,\beta$ have maximal
rank (in the present case $4k'$). In the sigma model context
\cite{Witten} this implies that all the $4k'$ right-handed chiral
fermions in $\Phi^{+Y'}_+$ are paired with a subset of $4k'$ left-handed ones
from $\Lambda^a_+$ and become massive. The sigma model described in this
section and first proposed by Witten \cite{Witten} corresponds to $k'$
instantons in $R^{4k}$ with gauge group $SO(n)$. The somewhat subtle point
about the allowed freedom of redefinition of the matrix $\Delta$ will be
discussed at the end of subsection \ref{oleg}.

It should be pointed out that throughout this subsection we always
preserved the $SU(2)'$ (indices $A'$) invariance of the free theory.
\footnote{The other half of the
automorphism group $SO(4)$ of $(0,4)$ supersymmetry (the group $SU(2)$,
indices $A$) is broken by the
presence of the constant matrices $\alpha^a_{AY'}$ in (\ref{lin}).} In
fact, this invariance {\it is not required
by supersymmetry}. Indeed, we can construct another potential term,
similar to (\ref{int}), but with broken $SU(2)'$. To this end we introduce
{\it manifest} $\theta^+$ {\it dependence}:
\begin{equation}\label{broken}
m\int d^2xdu d^2\theta^+_+\; \theta^{+A'}_+ M^{+a}_{A'}(X^+,u) \Lambda^a_+
\; .
\end{equation}
One might think that such a term would break supersymmetry. However, the
variation $\delta\theta^{+A'}_+ = \epsilon^{+A'}_+$ can be compensated by
a suitable transformation of the superfield $\Lambda$,
\begin{equation}\label{sus}
\delta \Lambda^a_+ = m\epsilon^{-A'}_+ M^{+a}_{A'}\; .
\end{equation}
Here we used $D^{++}\epsilon^{-A'}_+ = \epsilon^{+A'}_+$. Note that after
this $\Lambda^a_+$ ceases to be a superfield. In order for this trick to work
the matrix $M^{+a}_{A'}(X^+,u)$ must satisfy conditions similar to
(\ref{c1}),(\ref{c2}). These constraints imply a structure of
$M^{+a}_{A'}(X^+,u)$ like in (\ref{lin}), but this time involving constant
matrices with $SU(2)'$ indices. In a sense, the term (\ref{broken})
corresponds to introducing a {\it constant superfield}
$\Phi^{+A'}_+=\theta^{+A'}_+$ in (\ref{int}).
Yet another way to say this is that
we have replaced the field $\phi$ in (\ref{WZ}) by a unit matrix (and
put $\chi_-=0$). Note that the matrix $M^{+a}_{A'}(X^+,u)$ explicitly
breaks $SU(2)'$.
By analyzing the general restrictions on
the potential term following by $(0,4)$ supersymmetry, Witten
\cite{Witten} has also found such terms. He argued that they should not be
allowed if one wants to obtain a sigma model which makes contact with CFT.

To summarize this subsection, we have shown how $(0,4)$ supersymmetry can
efficiently determine the kind of interaction for the linear sigma model
and put it in correspondence with the starting point of the ADHM
constructions.  We emphasize the r\^ole of the abelian gauge invariance
(\ref{gau}) in fixing this interaction. Of course, the model we have found
is just the superspace version of the one of Witten.

\subsection{The instanton gauge field} \label{oleg}

In the ADHM constructions one is able, on the basis of the matrix $\Delta$
above, to give an explicit expression for the instanton gauge field. This can
also be done in our supersymmetric sigma model. One way is to follow Witten
and extract the gauge field from the component sigma model action. We shall
do this at the end of this subsection, but first we prefer to continue with
our manifestly supersymmetric approach. As we shall see, it will lead us
directly to the twistor counterpart of the SDYM field introduced in subsection
\ref{sdym}.

As we have shown in the beginning of subsection \ref{flat}, when $n=0$ and
the coupling is simply given by eq. (\ref{coup}), then the model describes
$4k'+4k'$ free massive bosons and fermions. To a certain extent this
remains true even for the interaction Lagrangian with (\ref{coup})
replaced by (\ref{int}). The more precise statement is that among the
$n+4k'$ left-handed fermions $\lambda_+$ contained in $\Lambda^a_+$ there
is a subset of $4k'$ which are paired with the right-handed fermions in
$\Phi$ and become massive (together with the bosons from $\Phi$). The
remaining chiral fermions stay massless. The question is how to separate
the massive from the massless modes, i.e., how to diagonalize the action.
We shall try a diagonalization which resembles (\ref{deco}).

As the first step let us complete the $2k'\times (n+4k')$ matrix
$v^{+a}_{Y'}(X^+,u)$ to a
full {\it orthogonal} matrix $v^{\hat aa}(X^+,u)$, where the $n+4k'$
dimensional index $\hat a = (+Y', -Y',i)$ and
$i=1,\ldots, n$ is an index of the group $SO(n)$. Orthogonality
means
\begin{equation}\label{or}
v^{\hat aa} v^{\hat b a} = \delta^{\hat a\hat b}\; ,
\end{equation}
where $\delta^{+Y', -Z'} = - \delta^{-Y', +Z'} = \epsilon^{Y'Z'}, \;\;
\delta^{+Y', +Z'}=\delta^{-Y', -Z'}=\delta^{\pm Y', i}=0 $.
In particular, for $\hat a = +Y'$ and $\hat b = +Z'$ we obtain just
condition (\ref{c1}). Since $v^{+a}_{Y'}$ is a function of $X^{+Y}$ and
$u^\pm$, we expect the other blocks of $v^{\hat aa}$, namely $v^{-Y'a}$
and $v^{ia}$ to be such functions too. Of course, the fact that
$v^{+a}_{Y'}$ must be a {\it linear} function of $X^{+Y}$ (see
(\ref{lin})) does not imply that the rest of $v^{\hat aa}$ are linear as
well. It is also clear that the new matrix blocks are not completely fixed
by (\ref{or}). The freedom consists of the subset of $SO(n+4k')$
transformations acting on the index $\hat a$ with analytic (i.e.,
functions of $X^{+Y}$ and $u^\pm$) parameters, which leave $v^{+a}_{Y'}$
invariant.

With the help of the newly introduced matrix we can make a change of variables
from the superfield $\Lambda^a_+$ to $\Lambda^{\hat a}_+ =v^{\hat
aa}\Lambda^a_+$. Then the gauge transformation (\ref{gauLv}) can be
translated into
\begin{equation}\label{gauL-}
\delta\Lambda^{-Y'}_+ = m\omega^{-Y'}_+, \ \ \ \delta\Lambda^{+Y'}_+ =
\delta\Lambda^{i}_+ = 0\; .
\end{equation}
As in the flat case of subsection \ref{flat}, eq. (\ref{gauL-}) allows us
to fix the supersymmetric gauge (cf. (\ref{sugau}))
\begin{equation}\label{sugauge}
\Lambda^{-Y'}_+ = 0\; .
\end{equation}
Now we can switch to the new superfields $\Lambda^{\hat a}_+$ in the kinetic
term for $\Lambda$ (\ref{acL}) and the coupling term (\ref{int}). It is
useful to introduce the following notation
\begin{equation}\label{VV}
(V^{++})^{\hat a\hat b} = v^{\hat aa}D^{++} v^{\hat ba}\; .
\end{equation}
Then the terms of the Lagrangian containing $\Lambda$ become
(in the gauge (\ref{sugauge}))
\begin{equation}\label{LLL}
{\cal L}^{++}_{++}(\Lambda) = {1\over 2}
\Lambda^i_+[\delta^{ij} D^{++} + (V^{++})^{ij}]
\Lambda^j_+  + \Lambda^{+Y'}_+[{1\over 2}(V)_{Y'Z'} \Lambda^{+Z'}_+ +
(V^{++})^{-i}_{Y'} \Lambda^{i}_+ + m\Phi^+_{+Y'}]\; ,
\end{equation}
where we have denoted $V_{Y'Z'} =(V^{++})^{--}_{Y'Z'}$.  We see that
$\Lambda^{+Y'}_+$ enters without derivatives, so it can be eliminated from
(\ref{LLL}). The result is
\begin{eqnarray}
{\cal L}^{++}_{++}(\Lambda)
&=& {1\over 2} \Lambda^i_+[\delta^{ij} D^{++} + (V^{++})^{ij}]  \Lambda^j_+
\nonumber \\
&-& \label{LL}
 {1\over 2}[(V^{++})^{-i}_{Y'} \Lambda^{i}_+ + m\Phi^+_{+Y'}] (V^{-1})^{Y'Z'}
[(V^{++})^{-j}_{Z'} \Lambda^{j}_+ + m\Phi^+_{+Z'}]\; .
\end{eqnarray}
The complete action of the model is obtained by adding the kinetic terms for
$X^+$ (\ref{acX}) and $\Phi^+_+$ (\ref{acP}).

To make contact with the free case (\ref{massac}) we can put the superfields
$\Lambda^i_+$ to 0 and $V_{Y'Z'} = \epsilon_{Y'Z'}$. What remains is just the
mass term for the superfields $\Phi$ (cf. (\ref{massac})). This explains
why the subset of superfields $\Lambda^{+Y'}_+$ (and, consequently, the
chiral fermions therein) correspond to massive modes. The more interesting
simplification occurs
 when we let the mass $m$ go to infinity. This corresponds to
suppressing the kinetic term for $\Phi$, after which the second term in
(\ref{LL}) becomes auxiliary and we can drop it. After that we are left with
the simple result
\begin{equation}\label{ADga}
{\cal L}^{++}_{++}(\Lambda)|_{m\rightarrow\infty} = \Lambda^i_+[\delta^{ij}
D^{++} +
({V}^{++})^{ij}]  \Lambda^j_+\; ,
\end{equation}
where
\begin{equation}\label{calV}
({ V}^{++})^{ij} = v^{ia}(X^+,u)D^{++}v^{ja}(X^+,u)\; .
\end{equation}
 Now, let us compare
this with the harmonic treatment of the SDYM equations, in particular,
with the covariant derivative $\nabla^{++}$ (\ref{hco}),(\ref{+++'}). We
realize that ${ V}^{++}$ in (\ref{ADga}) is the counterpart of the
holomorphic harmonic connection for the gauge group $SO(n)$ and
$\Lambda^i_+$ is the analog of the matter field. As explained in
subsection \ref{sdym}, starting form a given ${ V}^{++}(X^+,u)$ one
can reconstruct a solutions of the SDYM equations by first finding the
``pure gauge" transformation $h(X,u)$ from the equation $hD^{++}h^{-1} =
{ V}^{++}$ and then substituting it in (\ref{pg}). It is important to
realize that the matrices $h^{ij}(X^{AY}, u^\pm)$ belong to $SO(n)$ and
are {\it not} holomorphic (i.e., they depend on both $X^\pm$). In
(\ref{or}),(\ref{calV}) we have found another representation of
${V}^{++}(X^+,u)$, this time in terms of the {\it holomorphic}, but bigger
($SO(n+4k')$) matrices $v^{\hat aa}(X^+,u)$.
More precisely, ${V}^{++}(X^+,u)$ is given in terms of the rectangular
matrix $v^{ia}$, which is in turn determined by $v^{+a} \sim \Delta$.
O. Ogievetsky \cite{Oleg} has found another, direct expression for
$V^{++}$ in terms of $\Delta$ (at least for $R^4$
and the gauge group $SU(2)\sim Sp(1)$).  Starting from $\Delta$, he was
able to find the matrix $h^{ij}(X^{AY}, u^\pm)$ and from it to compute
${V}^{++}$. The result is remarkably simple. If we use the canonical
form of $\Delta$ (\ref{canon}), then
\begin{equation}\label{olegV}
({ V}^{++})^{ij} = b^{+i}_{Y'} (B^{-2})^{Y'Z'}b^{+j}_{Z'}
\end{equation}
with $B_{Y'Z'} = u^{+A}u^{-B} (B_{AY'BZ'} +X_{AB}\epsilon_{Y'Z'})$. It
would be very interesting to find out the correspondence between this
expression and ours (up to gauge transformations, of course).

To summarize, we have seen that the
manifestly supersymmetric formulation of our sigma model naturally makes
contact with the harmonic (or twistor) interpretation of self-duality.
On the other hand, we have an alternative way to extract the SDYM
field encoded in the action term (\ref{int}). It consists in obtaining the
part of the component action in which the massless chiral fermions
$\lambda_+$ are coupled to a composite gauge field. To this end we have to
use the {\it non-supersymmetric} Wess-Zumino gauge (\ref{WZ}) rather than
the manifestly supersymmetric one (\ref{sugauge}). To simplify our task we
shall keep only the relevant fields, i.e., the fermions in $\Phi^{+Y'}_+$ and
$\Lambda^a_+$ and the bosons in $X^+$. Since $\chi^{AY'}_-$ is
accompanied by $(\theta^+_+)^2$ in (\ref{WZ}),
the other superfields in (\ref{int}) contribute with their lowest order
components only,
i.e., $\Delta^a_{AY'}$ from (\ref{lin'}) and $\lambda^a_+$ from (\ref{Lam}).
This, together with the kinetic term for $\Lambda^a_+$ (\ref{comL}), gives
\begin{equation}\label{fer}
S = \int d^2xdu \; \left( {i\over 2}\lambda^{a}_+\partial_{--}\lambda^a_{+} +
i\sigma^{--a}_{-} \partial^{++}\lambda^a_+ + m u^-_A\chi^{AY'}_- u^+_B
\Delta^{Ba}_{Y'}(X)\lambda^a_+\right)\; .
\end{equation}
The important point here is that the Lagrange multiplier term with
$\sigma^{--a}_{-}$
has not changed, so we still have the field equation
$\partial^{++}\lambda^a_{+}(x,u) = 0 \
\rightarrow \ \lambda^a_{+} = \lambda^a_{+}(x)$, like in the free case. Then
the harmonic integral in (\ref{fer}) becomes trivial and we obtain the action
\begin{equation}\label{ferr}
S = \int d^2x \; \left( {i\over 2}\lambda^{a}_+\partial_{--}\lambda^a_{+}
 - {m\over 2} \chi^{AY'}_- \Delta^a_{AY'}(X)\lambda^a_+\right)\; .
\end{equation}
The subsequent steps are described by Witten and we only sketch them. One
introduces an $n\times (n+4k')$ matrix $v^a_i(X)$, orthogonal to $\Delta$,
$\Delta^a_{AY'}v^a_i=0$ and orthonormalized, $v^a_iv^a_j = \delta_{ij}$ (cf.
subsection \ref{quaternion}).
In a sense, this step is similar to the introduction of the matrix
$v^{\hat aa}$ (\ref{or}). The aim is to diagonalize (\ref{ferr}), i.e., to
separate the massive fermions from the massless ones. The massless
fermions are just $\lambda^i_+ = v^{ia}\lambda^a_+$.
After that one puts all massive fields to 0 (or, equivalently,
$m\rightarrow\infty$) and obtains
\begin{equation}\label{fergau}
S ={i\over 2} \int d^2x \; \lambda^i_+(\delta^{ij}\partial_{--} +
\partial_{--}X^{BY} A^{ij}_{BY})
\lambda^j_+\; ,
\end{equation}
where
\begin{equation}\label{ADHMgau}
A^{ij}_{BY} = v^{ia} {\partial v^{ja}\over \partial X^{BY}}\; .
\end{equation}
This is precisely the expression for the instanton field in the ADHM
construction, see (\ref{siex}).

In the context of comparing the sigma model to the ADHM
construction we would like to come back to the issue of counting the number of
instanton parameters.  As explained in subsection
\ref{quaternion},  the ADHM construction has
another important ingredient, namely the freedom to redefine the matrices
$\Delta$ and still obtain the same instanton solution. There one uses
$GL(k',Q)$ transformations of the index $Y'$ in the
defining constraints (\ref{ADHM2}), (\ref{orgo}). In the sigma model above the
kinetic term (\ref{acP}) is a quadratic form invariant under $Sp(k')$ only.
 The consequence is that there exist
equivalence classes of sigma models (obtained from one another by
transformations from $GL(k',Q)/Sp(k')$), which produce the same SDYM field.
However, the problem may not be so serious, as we only clearly see the SDYM
field in the limit $m\rightarrow\infty$,
where the kinetic term (\ref{acP}) is suppressed and, effectively, the
$GL(k',Q)$ invariance of the ADHM construction is restored.

In conclusion we can say that the supersymmetric sigma model described in
this section exhibits the following features. First, supersymmetry and the
gauge invariance (\ref{gau}) lead to the very special interaction
(\ref{int}). In it we found the matrix $\Delta$ (\ref{lin'}), which is the
starting point in the ADHM construction. Then, proceeding in a manifestly
supersymmetric way, we were able to separate the massless from the massive
modes in the model. In the massless part of the Lagrangian (\ref{ADga}) we
found the {\it harmonic connection} ${ V}^{++}$, which is the
``twistor transform" of the SDYM field. Alternatively, a
non-supersymmetric gauge allowed us to arrive at the {\it harmonic-free}
action (\ref{ferr}). Separating the massless and massive modes in it lead
to the ADHM instanton field, instead of its ``twistor transform".

\vskip5mm
{\bf Acknowledgements} The authors would like to thank E. Corrigan, F.
Delduc, Ch. Devchand, W. Nahm, O. Ogievetsky, G. Papadopoulos and E.
Witten for useful discussions. E.S. is
particularly indebted to F. E{\ss}ler, R. Flume and V. Rittenberg for
valuable suggestions and for the constant encouragement.

\newpage

\setcounter{equation}{0}
\noindent{\bf{\Large Appendix A}}
\def\theequation{A.\arabic{equation}}

\vskip5mm
Here we shall prove a ``no-go'' theorem about the off-shell content of the
$(0,4)$
chiral fermion supermultiplet. In this multiplet the only propagating
fields are chiral fermions $\lambda^a_+(x)$, with the action
\begin{equation}\label{a1}
S_{\lambda}= {i\over 2}\int d^2x \lambda^a_+\partial_{--}\lambda^a_+,
\end{equation}
where $a=1,2,\ldots, n,$ and $n$ is a positive integer.
Note that this action possesses an $SO(n)$ symmetry with $\lambda^a_+(x)$
transforming as $SO(n)$ vector.

Since on shell
$\partial_{--}\lambda^a_+=0$, the chiral fermions are inert under
$(0,4)$ supersymmetry. An off-shell multiplet of of chiral fermions must
also include  a number of auxiliary fields, since the number of fermionic
fields should match the  number of bosonic fields. The auxiliary fields
vanish on shell.

We assume the following:
\begin{itemize}
\item $(0,4)$ supersymmetry is realized linearly;
\item the multiplet fields form representations of $SO(n)$ and of the
supersymmetry automorphism group $SU(2)\times SU(2)'$;
\item the action is quadratic in the fields and invariant
under $SO(n)\times SU(2)\times SU(2)'$.
\end{itemize}
Given this, {\it there are no off-shell chiral fermion multiplets
with a {\bf finite} number of auxiliary fields, except for $n=4$, where
$SO(4)$ must be identified with $SU(2)\times SU(2)'$}.

To prove this, let us first consider an off-shell multiplet describing one real
chiral
fermion, $n=1$. Since the dimensions of $\lambda^a_+(x)$ and of the
supersymmetry parameter $\epsilon^{AA'}$ are half-integer,
the auxiliary bosonic fields should have integer dimensions $0, \pm 1, \pm 2,
\ldots$, and the auxiliary fermions should have half-integer dimensions.
Therefore, the auxiliary fermions must enter the Lagrangian (of dimension
2) in pairs, hence the total number of fermionic fields would be odd. Further,
the bosonic fields always occur in multiples of four: the supersymmetry
transformation implies that the bosons carry odd number of vector indices of
$SO(4)=SU(2)\times SU(2)'$ (decomposing such a representation into
irreducible ones, we find that all of them are multiples of four).  Hence
the fermionic and bosonic degrees of freedom do not match, and the off-shell
multiplet does not exist.

The above argument is immediately generalized to $n>1$. There
the auxiliary fields carry the same vector index $a$ of $SO(n)$
as the chiral fermions. This clearly leads to the same mismatch of degrees
of freedom. The only exception is the case $n=4$, where it is possible
to identify $SO(4)$ with $SU(2)\times SU(2)'$. \footnote{For $n=3$ $SO(3)$
can be identified with either $SU(2)$ or $SU(2)'$,
but this case is ruled out because the total number of fermions is odd.}
Now the four chiral fermions can be matched with four auxiliary
bosons \cite{Gates}. Note, however, that in such a formulation the $SO(4)$
symmetry of (\ref{a1}) is tied to the supersymmetry automorphism group. As a
consequence, it cannot be gauged independently, as required in the sigma model
of Witten.

We want to stress that the above argument is a counting one, i.e., based
on the assumption of {\it finite} numbers of auxiliary fields. As we have shown
in the text, off-shell representations with infinitely many auxiliary fields
exist for any number of chiral fermions.

\vskip10mm

\setcounter{equation}{0}
\noindent{\bf{\Large Appendix B}}
\def\theequation{B.\arabic{equation}}

\vskip5mm

The origin of the three types of $(0,4)$ superfields used in this paper
becomes more clear if we regard them as obtained by trivial dimensional
reduction and truncation from the two basic $N=2, D=4$ supermultiplets,
matter and gauge.

The adequate superspace for $N=2, D=4$ supersymmetry is again harmonic
 \cite{harms}.
The Grassmann variables in $N=2, D=4$ supersymmetry are two-component
$SL(2,C)$ spinors $\theta^A_\alpha$ and their complex conjugates
$\bar\theta^A_{\dot\alpha}$ with an additional $SU(2)$ automorphism index $A$.
The spinor covariant derivatives
\begin{equation}\label{D'}
D_{A\alpha} = {\partial\over\partial\theta^{A\alpha}} +
i\bar\theta_A^{\dot\alpha} {\partial\over\partial x^{\alpha\dot\alpha}}
\end{equation}
satisfy the $N=2, D=4$ supersymmetry algebra
\begin{equation}
\{D^\alpha_A, D^\beta_B\} = \{\bar D^{\dot\alpha}_A, \bar D^{\dot\beta}_B\}
= 0, \ \ \
\{D^\alpha_A,\bar D^{\dot\alpha}_B \} = 2i\epsilon_{AB}
\partial^{\alpha\dot\alpha}\; . \label{susy'}
\end{equation}
Projecting  the $SU(2)$ indices
$A,B$ with harmonic variables,
$D^{+\alpha} \equiv u^{+A}D^{\alpha}_A,\ \  \bar D^{+\dot\alpha} \equiv
u^{+A}\bar D^{\dot\alpha}_A$,
we can rewrite (\ref{susy'}) as follows
\begin{equation}\label{++'}
\{D^{+\alpha}, D^{+\beta}\} = \{\bar D^{+\dot\alpha}, \bar D^{+\dot\beta}\}
= \{D^{+\alpha},\bar D^{+\dot\beta}\}  =0\; .
\end{equation}
Grassmann analytic superfields are defined by the constraint
\begin{equation}\label{ga'}
D^{+\alpha}\Phi(x,\theta,u) = \bar D^{+\dot\alpha}\Phi(x,\theta,u) = 0\; ,
\end{equation}
that is explicitly solved in the analytic basis in harmonic superspace
\begin{equation}\label{bas'}
\hat x^{\alpha\dot\alpha} = x^{\alpha\dot\alpha} + 2i\theta^{A\alpha}
\bar\theta^{B\dot\alpha} u^+_{(A}u^-_{B)},
 \ \ \theta^{\pm\alpha}=u^\pm_A\theta^{A\alpha}, \ \
\bar\theta^{\pm\dot\alpha}=u^\pm_A\bar\theta^{A\dot\alpha} , \ \ u^\pm
\end{equation}
in the form $\Phi = \Phi(x, \theta^+, \bar\theta^+,u)$ (we shall
not write $\hat x$ explicitly). In this basis the spinor derivatives
$D^+$ are just partial derivatives, but the harmonic derivative $D^{++}$
becomes covariant:
\begin{equation}\label{harco}
D^{++} = \partial^{++} +2i\theta^{+\alpha}\bar\theta^{+\dot\alpha}
\partial_{\alpha\dot\alpha}\; .
\end{equation}

In $N=2, D=4$ rigid supersymmetry there are two basic supermultiplets, the
matter (hyper)multiplet and the gauge one (we do not discuss local
supersymmetry, i.e., supergravity). Both of them can be described by
appropriate analytic harmonic superfields \cite{harms}. Matter is
described in terms of a set of commuting superfields $q^{+a}(x, \theta^+,
\bar\theta^+,u)$ with the action
\begin{equation}\label{q+}
S_{q^+} = {1\over 2}\int d^4x du d^4\theta^+ \; q^{+a} D^{++} q^+_a\; ,
\end{equation}
where  the index $a$ corresponds to the $2n$ representation of $Sp(n)$.
In addition, these superfields satisfy the reality
condition $\widetilde{q^{+a}} = \epsilon_{ab}q^{+b}$.
It should be emphasized that off shell the superfields $q^+$ contain an
infinite number of auxiliary fields, in accord with the ``no-go" theorem of
ref. \cite{RocekSiegel}. Only on shell the equation of motion $D^{++}q^+ =
0$ eliminates the harmonic dependence and one obtains short multiplets
consisting of $4n$ real bosons and $4n$ fermions.
Here one should recall the similar equation (\ref{irr}), which was a
purely kinematical (off-shell) constraint in $N=(0,4), D=2$ supersymmetry.

The (abelian) gauge multiplet is described by a real commuting superfield of
the type $A^{++}$ undergoing the gauge transformation
$\delta A^{++} = D^{++}\Omega$
with an analytic superfield parameter $\Omega$. In the Wess-Zumino  gauge
the superfield
$A^{++}$ has the short harmonic expansion
\begin{eqnarray}
A^{++} &=& [(\theta^+)^2 M(x) + c.c.]
+ i\theta^{+\alpha}\bar\theta^{+\dot\alpha} A_{\alpha\dot\alpha}(x) +
[(\bar\theta^+)^2 \theta^{+\alpha}u^-_A \chi^{A}_\alpha(x) + c.c.]
\nonumber \\
&+& (\theta^+)^2(\bar\theta^+)^2 u^-_Au^-_B D^{AB}(x)\; . \label{WZ'}
\end{eqnarray}
Here $M$ is the physical scalar (complex), $A$ is the gauge field,
$\chi$ are the physical spinors and $D$ are auxiliary fields. Note that this
gauge multiplet, unlike the matter one above, does not need infinitely many
auxiliary fields off shell.

The action for $A^{++}$ resembles (\ref{acP}):
\begin{equation}\label{acP'}
S_A = \int d^4x d^8\theta du_1du_2 {1\over (u^+_1u^+_2)^2} A^{++}(1)
A^{++}(2)\; .
\end{equation}
This time we had to use the harmonic distribution $(u^+_1u^+_2)^{-2}$ in
order to match the $U(1)$ charges. Instead of (\ref{df}), it satisfies the
identity
\begin{equation}\label{df'}
D^{++}_1{1\over (u^+_1u^+_2)^2} = D^{--}_1\delta^{+2,-2}(u_1,u_2)\; ,
\end{equation}
which is still sufficient to prove the gauge invariance of (\ref{acP'}).

Now we dimensionally reduce the above superfields to $D=2$. The resulting
$(4,4)$ superfields depend on the left- and right-handed
Grassmann variables  $\theta^{+A'}_{\pm}$. Further, they can be truncated
to $(0,4)$ superfields. This simply means that we decompose the $(4,4)$
superfields in the right-handed $\theta^+_-$ and then discard some of the
resulting $(0,4)$ superfields. Thus, the abelian gauge superfield has the
decomposition
\begin{equation}\label{decA}
A^{++} = B^{++} + \theta^{+A'}_-\Phi^{+}_{+A'} +
(\theta^+_-)^2C_{++}\; .
\end{equation}
Each of the coefficients here is a $(0,4)$ superfield, $B$ and $C$ are even,
whereas $\Phi$ is odd. It is quite obvious that $\Phi^{+A'}_+$ is the
model for the odd abelian gauge superfield (\ref{Phi}). The decomposition
and truncation of the gauge transformation law for $A^{++}$
explain (\ref{gau}). A similar procedure leads from the gauge invariant
action (\ref{acP'}) to its $(0,4)$ counterpart (\ref{acP}) (a factor of
$u^+_1u^+_2$ appears in the numerator due to the integration over
$\theta^+_-$).

The other two multiplets used in section \ref{04} both have their origin in
an $N=2, D=4$ hypermultiplet, they just correspond to different truncations.
Thus, from the decomposition of $q^{+a}$ we take only the odd term
\begin{equation}\label{decq}
q^{+a} = \theta^{+}_{-A'} \Lambda_+^{A'a}
\end{equation}
and from that of $X^{+Y}$ only the even terms
\begin{equation}\label{decx}
X^{+Y} = X^{+Y} +(\theta^+_-)^2 P^{-Y}_{++}\; .
\end{equation}
In both cases the corresponding actions (\ref{acL}) and (\ref{acX}) are
obtained from the hypermultiplet action (\ref{q+}) (the superfield
$P^{-Y}_{++}$ serves as a Lagrange multiplier in the action for $X^+$,
which produces the condition (\ref{irr})).

We should point out that the above truncation gives the correct type of
$(0,4)$ superfields, but does not completely reproduce the symmetry of the
$(0,4)$ sigma model. For instance, the superfield $\Lambda_+^{A'a}$ in
(\ref{decq}) describes the chiral
fermions in multiples of 4, but we know this
does not have be the case in the $(0,4)$ sigma model. The reason is that
$(0,4)$ supersymmetry is less restrictive than the $(4,4)$ one.

\end{document}